\documentclass[twocolumn]{aastex61}

\received{\today}
\submitjournal{ApJ}

\shorttitle{Ethanol tree}
\shortauthors{Skouteris et al.}

\begin{document}

\title{The genealogical tree of ethanol: gas-phase formation of
  glycolaldehyde, acetic acid and formic acid}

\correspondingauthor{D. Skouteris, N. Balucani, C. Ceccarelli}
\email{dimitrios.skouteris@sns.it, nadia.balucani@unipg.it, cecilia.ceccarelli@univ-grenoble-alpes.fr}

\author{Dimitrios Skouteris}
\affil{Scuola Normale Superiore, Piazza dei Cavalieri 7, 56126 Pisa, Italy}

\author{Nadia Balucani}
\affil{Dipartimento di Chimica, Biologia e Biotecnologie, Universit\`a degli Studi di Perugia, Via Elce di Sotto 8, 06123 Perugia, Italy}
\affil{Univ. Grenoble Alpes, CNRS, IPAG, F-38000 Grenoble, France}
\affil{INAF-Osservatorio Astrofisico di Arcetri, Largo E. Fermi 5, I-50125, Florence, Italy}

\author{Cecilia Ceccarelli}
\affil{Univ. Grenoble Alpes, CNRS, IPAG, F-38000 Grenoble, France}

\author{Fanny Vazart}
\affil{Scuola Normale Superiore, Piazza dei Cavalieri 7, 56126 Pisa, Italy}

\author{Cristina Puzzarini}
\affil{Dipartimento di Chimica "Giacomo Ciamician", Universit\`a di Bologna, Via F. Selmi 2, 40126 Bologna, Italy}
\affil{INAF-Osservatorio Astrofisico di Arcetri, Largo E. Fermi 5, I-50125, Florence, Italy}

\author{Vincenzo Barone}
\affil{Scuola Normale Superiore, Piazza dei Cavalieri 7, 56126 Pisa, Italy}

\author{Claudio Codella}
\affil{INAF-Osservatorio Astrofisico di Arcetri, Largo E. Fermi 5, I-50125, Florence, Italy}

\author{Bertrand Lefloch}
\affil{Univ. Grenoble Alpes, CNRS, IPAG, F-38000 Grenoble, France}

\begin{abstract} 
  Despite the harsh conditions of the interstellar medium, chemistry thrives in it,
  especially in star forming regions where several interstellar
  complex organic molecules (iCOMs) have been detected. Yet, how these species are synthesised is a
  mystery. The majority of current models claim that this happens on
  interstellar grain surfaces. Nevertheless, evidence is mounting
  that neutral gas-phase chemistry plays an important role. In
  this article, we propose a new scheme for the gas-phase synthesis of
  glycolaldehyde, a species with a prebiotic potential and for which
  no gas-phase formation route was previously known. In the proposed
  scheme, the ancestor is ethanol and the glycolaldehyde sister
  species are acetic acid (another iCOM with unknown
  gas-phase formation routes) and formic acid.  For the reactions of
  the new scheme with no available data, we have performed electronic
  structure and kinetics calculations deriving rate coefficients and
  branching ratios. Furthermore, after a careful review of the
  chemistry literature, we revised the available chemical networks, adding and
  correcting several reactions related to glycolaldehyde, acetic acid
  and formic acid. The new chemical network has been used in an
  astrochemical model to predict the abundance of glycolaldehyde,
  acetic acid and formic acid. The predicted abundance of
  glycolaldehyde depends on the ethanol abundance in the gas phase and
  is in excellent agreement with the measured one in hot corinos and
  shock sites. Our new model overpredicts the
  abundance of acetic acid and formic acid by about a factor of ten,
  which might imply a yet incomplete reaction network.

\end{abstract}



\section{Introduction}

About one third of the $ca.$ 200 molecules detected in the
Inter-Stellar Medium (ISM) are constituted by six or more atoms
(source {\it http://www.astro.uni-koeln.de/cdms/molecules}). All these
molecules contain at least one carbon atom. In the following, we will call iCOMs \footnote{Please note that we added “i” to the commonly used COMs acronym in order to be clear that these molecules are only complex in the interstellar context (Ceccarelli et al. 2017), contrary to what chemists would consider complex in the terrestrial context.}
(for interstellar Complex Organic Molecules) molecules with at least
six atoms and containing at least one carbon atom. In the majority of
iCOMs, hydrogen and oxygen are the additional elements. To date, slightly
more than forty iCOMs which contain elements other than C and H have been
detected. Thus, they represent about 20\% of detected ISM molecules.

Even though the presence of iCOMs has been known for decades (for
instance formamide has been detected in 1971 by Rubin et al.), the
processes that lead to their synthesis are still hotly
debated. Specifically, it is nowadays often assumed that iCOMs are
mostly synthesised on grain surfaces during the so-called warm-up
phase, when various radicals trapped in the grain mantles acquire
mobility and recombine into large molecules (e.g. Garrod \& Herbst
2006; Garrod et al. 2008). Yet, recent detections of iCOMs in cold
environments (Bacmann et al. 2012; Cernicharo et a. 2012; Jaber et
al. 2014; Vastel et al. 2014; Jim\'enez-Serra et al. 2016) have
challenged this exclusive role of grain-surface chemistry. Clearly,
some gas-phase chemistry is at work in cold environments
(Vasyunin \& Herbst 2013; Balucani et al. 2015; Ruaud et al. 2016;
Vasyunin et al. 2017).

Supporting the idea that grain-surface chemistry cannot synthesise all
detected iCOMs, recent quantum chemistry calculations have
shown that the combination of radicals trapped in amorphous water ice
does not necessarily lead to larger molecules (Enrique-Romero et
al. 2016; Rimola et al. 2017), in particular to iCOMs, 
as predicted by the above mentioned grain-surface chemical models. 
The basic reason is that radicals are
not oriented in a way for the recombination reaction to occur, as they
are trapped by the water-ice molecules in a configuration that 
favors other two-product reactions.

Following up with the idea that gas-phase reactions might have been
overlooked, Kahane et al. (2013) proposed that formamide (NH$_2$CHO)
is formed by the reaction of formaldehyde (H$_2$CO) and amidogen
(NH$_2$). Barone et al. (2015), Vazart et al. (2016) and Skouteris et
al. (2017) carried out theoretical computations showing that this
reaction can efficiently occur at low temperatures (contrarily to what claimed by
Song \& Kastner, 2016) and can explain the available
observations. Other studies have explored possible gas-phase
ion-neutral reactions leading to formamide (Spezia et al. 2016).
More recently, an observational study obtained with the IRAM-NOEMA
interferometer provided additional support to the gas-phase formation
of formamide (Codella et al. 2017).  Similarly, observations of its
deuterated forms by Coutens et al. (2016) are in good agreement with
the theoretical predictions of deuterated amidogen or formaldehyde
leading to deuterated formamide (Skouteris et al. 2017).

In summary, it is well possible that gas-phase reactions play an
important role in the formation of iCOMs, but more systems need to be
studied to understand their real importance.

The aim of this paper is to understand whether previously overlooked
gas phase routes can lead to glycolaldehyde (HCOCH$_2$OH), a
``special'' iCOM because of its prebiotic potential. Glycolaldehyde
has been detected towards the giant molecular cloud complex SgrB2
(Hollis et al. 2000) towards high- and low- mass star forming regions
(Beltran et al. 2009; Jorgensen et al. 2012, 2016; Coutens et al.
2015; Taquet et al. 2015; De Simone et al. 2017) and in shocked
regions (Lefloch et al. 2017). Several mechanisms of glycolaldehyde
synthesis on grain surfaces were proposed: they involve
recombination of radicals (Garrod et al. 2008), UV or particle irradiation (Woods et
al. 2012; Maity et al. 2014; Fedoseev et al. 2015; Butscher et
al. 2015, 2016, Chuang et al. 2017). 
On the contrary, previous work on glycolaldehyde formation in the gas
phase could not identify plausible interstellar routes (Wang \& Bowie
2010; Jalbout 2007).

In a recent work, Lefloch et al. (2017) showed that there is a
correlation between the abundances of glycolaldehyde and ethanol, even
though this is based on only four sources. Following the suggestion
provided by this possible correlation, in this work we propose a
series of gas-phase reactions that start from ethanol (CH$_3$CH$_2$OH)
and that lead to the synthesis of glycolaldehyde in a sequence of
gas-phase reactions which is similar to that connecting dimethyl ether
and methyl formate, two other common iCOMs (Balucani et al. 2015). As
in that case, the sequence of reactions starts with the
conversion of ethanol (or its isomer dimethyl ether in the case
analyzed by Balucani et al. 2015) to a reactive radical that can
further react with abundant O atoms leading to glycolaldehyde 
(methyl formate in the case analyzed by Balucani et al. 2015). Some of
the necessary data to test this hypothesis were available in the
literature (see Sec. 3). Some crucial data were instead missing. Therefore,
to verify whether the proposed route is efficient in the ISM
conditions, we have performed dedicated electronic structure and
kinetics calculations for the missing reactions. The main result of
this work is that ethanol can be considered not only the ancestor
of glycolaldehyde, but also of formic acid (HCOOH) and acetic acid
(CH$_3$COOH), another common iCOM. In addition, ethanol is revealed
to be one of the precursors of acetaldehyde, a widely spread iCOM. In other words,
ethanol can be considered the progenitor of three iCOMs and of formic acid.

The article is organised as follows.  
In Section
\ref{sec:old-reactions-acoms}, we briefly review the previously known
gas-phase reactions of the above iCOMs and formic acid. We then present the
overall scheme and justification of the newly proposed reactions in Section
\ref{sec:new-reaction-scheme} and the employed methodology and results of
our computations in Section \ref{sec:comp-deta-results}. In
Section \ref{sec:astrochemical-model} we present the predictions
obtained by an astrochemical model including the new reactions and we
discuss the comparison with observations. Section \ref{sec:disc-concl}
concludes this article.


\section{Previous gas-phase reactions leading to glycolaldehyde, acetic acid, formic acid 
and acetaldehyde}\label{sec:old-reactions-acoms}
In this section, we briefly review the gas-phase reactions forming
glycolaldehyde, acetic acid, formic acid and acetaldehyde, listed in
the publicly available chemical databases, KIDA ({\it
  http://kida.obs.u-bordeaux1.fr:}  Wakelam et al. 2015), UMIST
({\it http://udfa.ajmarkwick.net:} McElroy et al. 2013), and in the literature.

\subsection{Glycolaldehyde (HCOCH$_2$OH)} 
No reactions are reported in the KIDA or UMIST databases.
Halfen et al. (2006) proposed that protonated formaldehyde
(H$_2$COH$^+$) could react with formaldehyde to produce protonated
glycolaldehyde in a radiative association reaction. The electron
recombination of protonated glycolaldehyde then ends in glycolaldehyde
by losing an H atom. However, electronic structure calculations by
Horn et al. (2004) showed that the relative association products do
not have the molecular structure of protonated glycolaldehyde.
Furthermore, it has been known since the experimental work by Karpas \& Klein (1975) 
that two-product exothermic channels are available for the H$_2$COH$^+$ + H$_2$COH reaction, which
strongly reduces the probability of radiative association in the absence of secondary collisions. Finally,
Woods et al. (2012, 2013) claimed that this route is inefficient and
cannot reproduce the observed abundances.

\subsection{Acetic acid (CH$_3$COOH)} 
No reactions are reported in the KIDA or UMIST databases for this species and we are
not aware of proposed schemes of its formation in the gas phase.

\subsection{Formic acid (HCOOH)} 
KIDA lists one reaction, CH$_3$O$_2^+$ + e$^-$, which is assumed
to produce 50\% of HCOOH (+ H) and 50\% of CO$_2$ (+ H$_2$ + H). In
UMIST, the same reaction, reported as HCOOH$_2^+$ + e$^-$, is globally
faster, but has a branching ratio of only 13\% for the HCOOH channel,
with the major channel leading to HCO + OH + H. The UMIST rate coefficients are
based on the experiments by Vigren et al. (2013), who were, however,
only able to demonstrate that heavy products with at least one C and
two O atoms account for 13\% of the global reaction. This  could
include also CO$_2$ formation, as suggested in KIDA. In addition to
that, there are issues concerning the formation of the  HCOOH$_2^+$ /
CH$_3$O$_2^+$  isomers. The main formation route of so-called protonated
formic acid is considered to be the radiative association reaction
HCO$^+$ + H$_2$O (Herbst 1985). In the KIDA and UMIST networks, this
process is present with a relatively high rate coefficient of $1.7\times
10^{-12}$ cm$^3$ s$^{-1}$ at 100 K. In general, the rate coefficients of most radiative association
reactions are poorly defined and can only be estimated. From what is
known so far, a significant probability for radiative association
reactions to occur can be expected only when there are no exothermic
two-product channels or when the presence of very high exit barriers
prevents a fast escape from the potential well associated to the
addition intermediate. Only in these cases, indeed, the lifetime of
the intermediate can be long enough to permit the spontaneous emission
of photons necessary to its stabilization (in the absence of ternary
collisions, as in interstellar environments). As warned by Herbst
(1985), this is not the case of the HCO$^+$ + H$_2$O which is indeed a
fast reaction ($3.64\times10^{-9}$ cm$^3$ s$^{-1}$ at 100 K) with a very exothermic two-product
channel (leading to CO + H$_3$O$^+$). For this reason, we
have deleted the radiative association of HCO$^+$ + H$_2$O from our
reaction network. The other routes of HCOOH$_2^+$ / CH$_3$O$_2^+$ formation have
already been proved to be marginal (Vigren et al. 2013). In addition,
UMIST also reports OH + H$_2$CO $\rightarrow$ HCOOH + H, a reaction
which has been widely studied at higher temperatures. Since the channel
leading to HCOOH + H has an entrance barrier of 23.8 kJ/mol (Xu \& Lin 2007), we did not consider it in our network. Other reactions
listed in UMIST are expected to make a negligible contribution to HCOOH formation.

\subsection{Acetaldehyde (CH$_3$CHO)} 
Several reactions forming acetaldehyde are reported in KIDA. A
possibly major formation route is the electron recombination of protonated
acetaldehyde which ends up in acetaldehyde. However, the protonated
acetaldehyde is mostly formed by the reaction CH$_3$OCH$_3$ +
H$^+~~ \rightarrow$ CH$_3$CHOH$^+$ + H$_2$, which would imply a
substantial and improbable rearrangement of the nuclei. We dropped,
therefore, this reaction from the network. Another major formation
reaction route, and often the most important one in several published
models, involves atomic oxygen and ethyl radical: CH$_3$CH$_2$ + O
$\rightarrow$ CH$_3$CHO + H (Charnley et al. 2004; Codella et al.
2015; Vastel et al. 2014).

\section{New reaction scheme}\label{sec:new-reaction-scheme}

Figure \ref{fig:new-scheme} illustrates the scheme of the reactions
proposed in this work. 

In the proposed new reaction scheme, ethanol is chemically activated
by one of the abundant atomic or molecular radicals that are present
in interstellar clouds.  In Balucani et al. (2015), several radicals
were considered and atomic chlorine was suggested as the major
contributor in converting dimethyl ether into the reactive
methylmethoxy radical.  Vasyunin et al. (2017) considered, instead,
that the abundant OH radicals are the major players by referring to
the recent work by Shannon et al. (2014), who performed kinetics
experiments at T as low as 60 K\footnote{We would like to note, however, that
  the most appropriate value for the rate coefficient of the bimolecular reaction leading to
  CH$_3$OCH$_2$ + H$_2$O is about one half of that quoted by Vasyunin
  et al. (2017) because of the pressure dependence noted and discussed
  by Shannon et al. (2014) in their work.}.

We follow the same approach here, that is, we have considered the
reaction of ethanol with Cl atoms and OH radicals as the initiating
steps. In the following we give details of the employed reactions.

\subsection{The initiating reaction Cl + CH$_3$CH$_2$OH} 
This reaction has been widely investigated at room or higher
temperatures. In particular, Taatjes et al. (1999) were able to
derive the H-abstraction site-specific rate coefficients as,
differently from the case of dimethyl ether, there are three different
kinds of hydrogen atoms that Cl (or other radicals) can abstract: (i)
three equivalent H atoms from the methyl group (CH$_3$), (ii) two
equivalent H atoms from the methylene group (CH$_2$) and (iii) one H
atom from the hydroxyl group (OH). 
According to the measurements by Taatjes et al. (1999) at room temperature, H abstraction
from the methylene group (ii) is by far the dominant pathway
accounting for about 90\% of the total reaction.  H abstraction from
the methyl group (i) accounts for the rest, while abstraction of the
hydroxyl hydrogen (iii) is negligible.  Unfortunately, there are no
experimental data at the low temperatures of interest in our
case. Therefore, in our network we have included the reaction channels
(labelled reaction 1 and 2 in Table \ref{tab:our-new-reacs}) with
their room temperature values. In addition, the product branching ratio for the
reaction Cl + CH$_3$CH$_2$OH could slightly vary with the temperature as
channel (1) is exothermic
by 44.8 kJ/mol, channel (2) by 17.6 kJ/mol, while the channel
leading to CH$_3$CH$_2$O is slightly endothermic by 3.8 kJ/mol
(Rusic et al. 1994). Moreover, electronic structure calculations by
Rudic et al. (2002) predicted a very small barrier of \it ca. \rm 3 kJ/mol
for channel (2). 

\subsection{The initiating reaction OH + CH$_3$CH$_2$OH} 
This reaction has been widely investigated at room or higher
temperatures and, more recently, also at temperatures as low as 50 K
by Caravan et al. (2015).  Similarly to the case of the analogous
reaction with dimethyl ether in low temperature experiments, the
observed pressure dependence of the rate coefficients provided
evidence that, in addition to the bimolecular abstraction channel
leading to products, collisional stabilization of the weakly bound
OH-ethanol complex occurred under their experimental conditions $-$
thus providing an artificially high rate coefficient (we remind that
such a mechanism cannot be present under the rarefied conditions of
the interstellar medium). In our network, we have considered the value
of the rate coefficient that Caravan et al. (2015) recommended as
representative of the sole two-product channel, that is, at 82--91 K,
2.7 ($\pm$ 0.8)$\times 10^{-11}$ cm$^3$s$^{-1}$.  Also in this case
the H-abstraction can occur at three different sites, leading to the
radicals CH$_2$CH$_2$OH, CH$_3$CHOH and CH$_3$CH$_2$O.  As CH$_3$O (+
H$_2$O) was determined to be the major product in the analogous
reaction OH + CH$_3$OH by Shannon et al. (2013), Caravan et al. (2015)
attempted the detection of CH$_3$CH$_2$O, but failed. At higher T, the
formation of CH$_3$CHOH (+ H$_2$O) is known to be the main channel,
with a branching ratio varying between 0.75$-$0.9, while the
CH$_2$CH$_2$OH (+ H$_2$O) channel accounts for the rest (Marinov, 1999; Carr et al. 2011).  The
branching ratio was also seen to vary with T. We have, therefore,
decided to test two different scenarios: in the first scenario, we
have assumed a value of 0.9:0.1 (reactions 3a and 4a in Table 1); in
the second scenario, we have assumed a value of 0.7:0.3 (reactions 3b
and 4b in Table 1).  Even though these seem to be reasonable ranges, a
final value could be adopted only when low T determination of the
branching ratio becomes available.

\subsection{Second step: reactions of CH$_3$CHOH and CH$_2$CH$_2$OH with O}

\begin{figure}
  \plotone{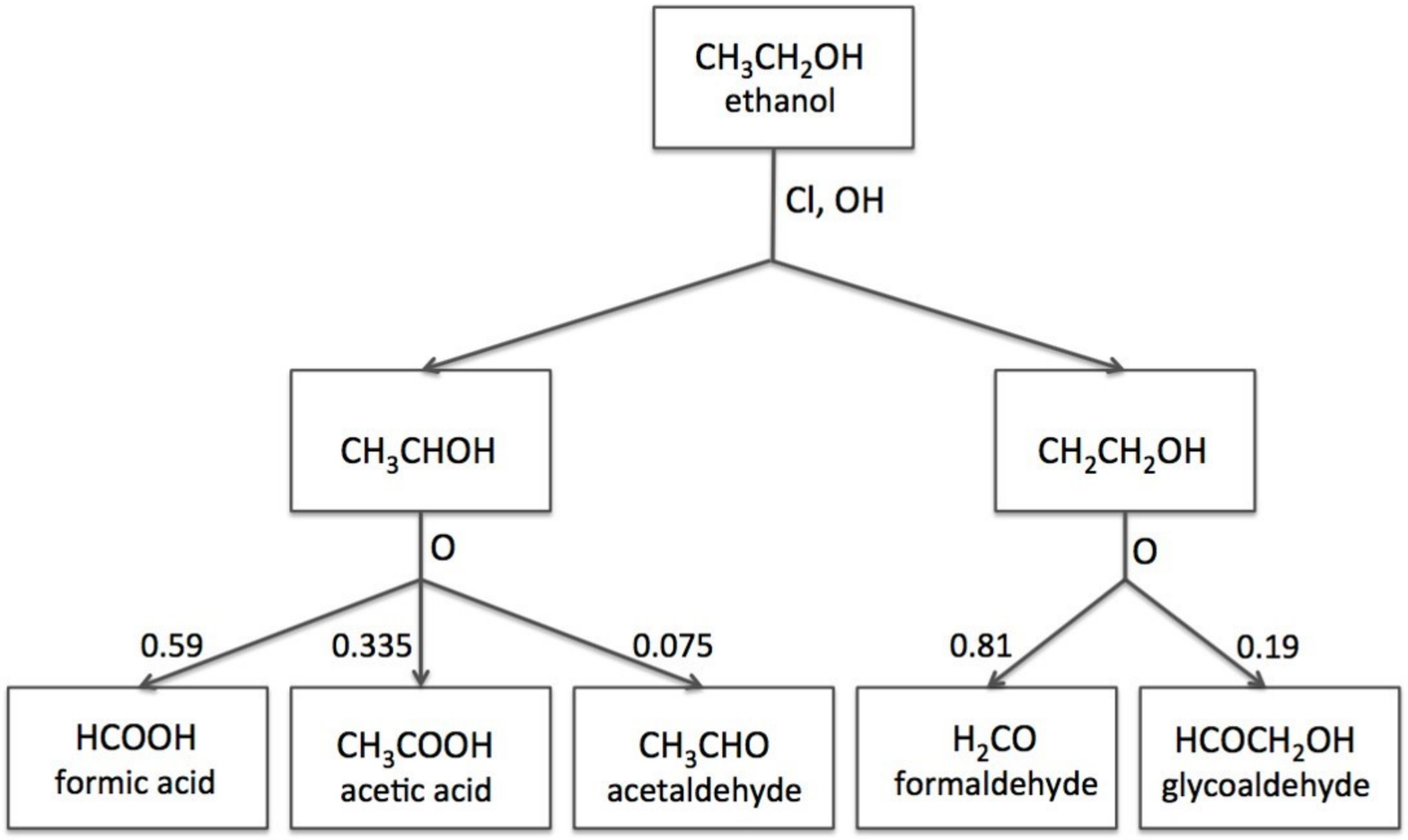}
  \caption{Scheme of the reactions starting from ethanol. The numbers
    indicate the relative branching ratios (see text).}
  \label{fig:new-scheme}
\end{figure}

Further reactions of the radicals produced in the initiating steps
with atomic oxygen generate the species shown in
Fig. \ref{fig:new-scheme}. To the best of our knowledge, only
fragmentary data were available in the literature concerning the
reaction channel O + CH$_3$CHOH $\rightarrow$ CH$_3$CHO + OH
(Edelbuttel-Einhaus et al., 1992) , so we have performed dedicated
electronic structure and kinetics calculations. The results of the
calculations are reported in Section 4.

The resulting overall picure is that CH$_3$CHOH and CH$_2$CH$_2$OH can form: 
(i) starting from CH$_3$CHOH
formic acid (59\%), acetaldehyde (7.5\%), acetic acid (33.5\%), and
(ii) starting from CH$_2$CH$_2$OH glycolaldehyde (19\%) and H$_2$CO (81\%),
respectively.

\begin{table*}
  \centering
  \begin{tabular}{lclccccc}
    \hline
    Reaction &&& $\alpha$ & $\beta$ & $\gamma$ & Label & Notes\\
    \hline
    CH$_3$CH$_2$OH + Cl &$\rightarrow$& CH$_3$CHOH + HCl        & 9.8(-11) & 0 & 0 & 1 & 1\\
    CH$_3$CH$_2$OH + Cl &$\rightarrow$& CH$_2$CH$_2$OH + HCl & 1.1(-11) & 0 & 0 & 2 & 1\\
    CH$_3$CH$_2$OH + OH &$\rightarrow$& CH$_3$CHOH + H$_2$O         & 2.4(-11) & 0 & 0 & 3a & 2\\
    CH$_3$CH$_2$OH + OH &$\rightarrow$& CH$_3$CHOH + H$_2$O         & 1.9(-11) & 0 & 0 & 3b & 2\\
    CH$_3$CH$_2$OH + OH &$\rightarrow$& CH$_2$CH$_2$OH + H$_2$O  & 2.7(-12) & 0 & 0 & 4a & 2\\
    CH$_3$CH$_2$OH + OH &$\rightarrow$& CH$_2$CH$_2$OH + H$_2$O  & 8.1(-12) & 0 & 0 & 4b & 2\\
    CH$_3$CHOH + O        &$\rightarrow$& HCOOH + CH$_3$          & 3.9(-10) & 0.18 & 0.49 & 5 & 3\\
    CH$_3$CHOH + O        &$\rightarrow$& CH$_3$CHO + OH          & 4.8(-11) & 0.19 & 0.39 & 6 & 3\\
    CH$_3$CHOH + O        &$\rightarrow$& CH$_3$COOH + H          & 2.2(-10) & 0.16 & 0.59 & 7 & 3\\
    CH$_2$CH$_2$OH + O &$\rightarrow$& HCOCH$_2$OH + H        & 1.1(-10) & 0.16 & 0.55 & 8 & 3\\
    CH$_2$CH$_2$OH + O &$\rightarrow$& H$_2$CO + CH$_2$OH   & 4.6(-10) & 0.17 & 0.51 & 9 & 3\\
    \hline
  \end{tabular}
  \caption{List of the reactions of the proposed scheme to form
    glycolaldehyde and acetic acid from ethanol.  $\alpha$, $\beta$
    and $\gamma$ are the coefficients for the rate constants, computed
    according to the usual
    equation $k = \alpha \times (T_{gas}/300K)^\beta \times
    exp[-\gamma/T_{gas}$]. The last two columns report the reaction labels 
    and notes. 
    $^1$ 
    We have used the (rounded) values at 300 K measured by Taatjes
    et al. (1999).
    $^2$  We have adopted the total value measured by Caravan et al (2015)
    in the range 82$-$91 K, 
    partitioned according to two possible scenarios for the channels 3 and 4 (see text).
     $^3$ The rate coefficients and product
    branching ratios are those computed in the present work. }
  \label{tab:our-new-reacs}
\end{table*}

The reactions with their branching ratios and rate coefficients are reported in
Table \ref{tab:our-new-reacs}.

\section{Computational details and results}\label{sec:comp-deta-results}

\subsection{Electronic structure calculations}
\label{sec:electronic} 

Calculations have been performed with a development version of the
Gaussian suite of programs (Frisch et al. 2013) as well as with the CFOUR program package. \footnote{(CFOUR, a quantum chemical
program package written by Stanton, J. F., Gauss, J., Harding, M. E., Szalay, P. G. with contributions from 
Auer, A. A., Bartlett, R. J., Benedikt, U., et al. and the integral packages MOLECULE (Alml\"of, J., \& Taylor, P. R.), PROPS 
(Taylor, P. R.), ABACUS (Helgaker, T., Jensen, H. J. Aa., J\o rgensen, P., \& Olsen, J.) and ECP routines by Mitin, A. V., 
\& van W\"ullen C. For the current version, see http://www.cfour.de).} Geometry optimizations for all stationary points were performed with the double-hybrid B2PLYP functional
(Grimme 2006) in conjunction with the m-aug-cc-pVTZ basis set
(Papajak et al., 2009; Dunning 1989) where $d$ functions on
hydrogens have been removed. Semiempirical dispersion contributions
were also included by means of the D3BJ model of Grimme (Goerigk \&
Grimme 2011; Grimme et al. 2011). Full geometry optimizations
have been performed for all molecules checking the nature of the
obtained structures (minima or first order saddle points) by
diagonalizing their Hessians. For each stationary points, the anharmonic force
field has been computed at the B2PLYP-D3BJ/m-aug-cc-pVTZ
level in order to evaluate the zero-point energies (ZPEs) using 
vibrational perturbation theory (VPT2). To obtain accurate electronic energies, the coupled-cluster singles and doubles approximation augmented by a 
perturbative treatment of triple excitations (CCSD(T), Raghavachari et al. 1989)
was employed in conjunction with extrapolation to the complete basis set limit (CBS) and inclusion of core-correlation effects (CV), thus leading to the so-called CCSD(T)/CBS+CV 
approach (Heckert et al. 2005, Heckert et al. 2006). The cc-pVnZ, with n=T,Q, basis sets (Dunning 1989) were used in the 
extrapolation to the CBS limit, while the cc-pCVTZ set (Woon \& Dunning 1995) was employed for evaluating the CV correction.

\subsection{Reaction paths for the reactions O + CH$_2$CH$_2$OH  and O + CH$_3$CHOH}

\begin{figure*}
  \plotone{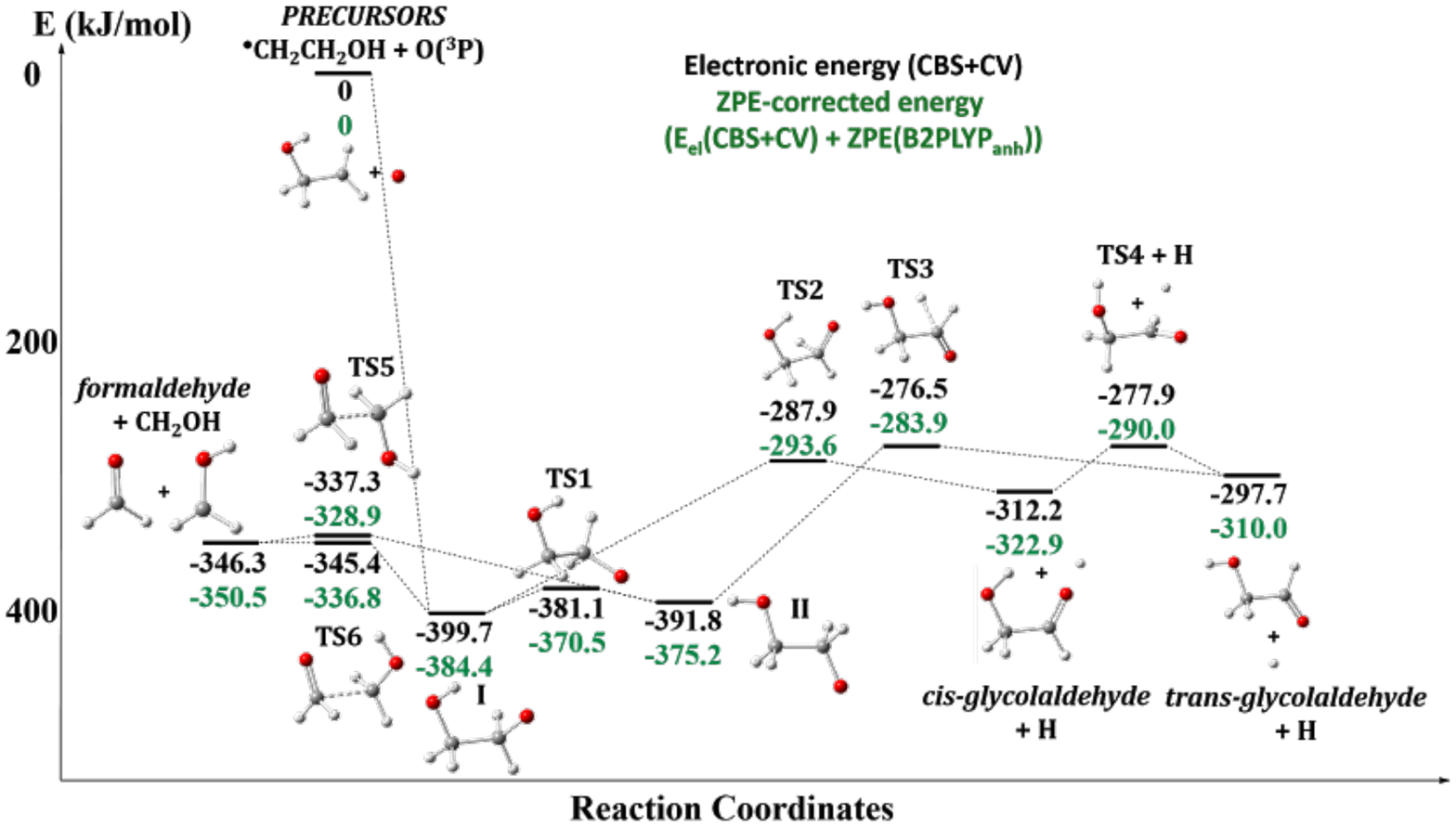}
  \caption{Reaction path for the O($^3$P) + CH$_2$CH$_2$OH scheme. Both electronic (above) and zero-point corrected (below) energies are shown.}
  \label{fig:glyc}
\end{figure*}

\begin{figure*}
  \plotone{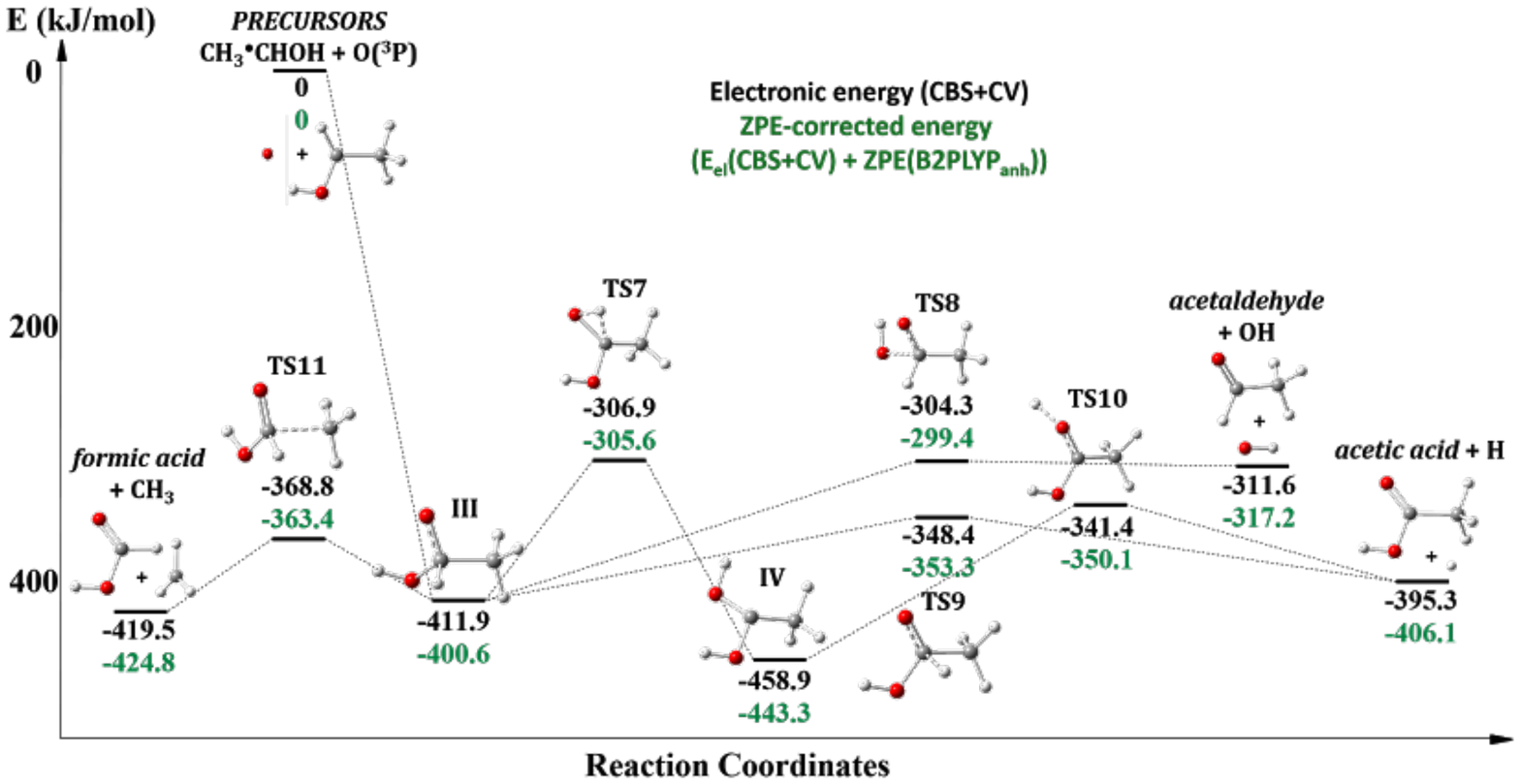}
  \caption{Reaction path for the O($^3$P) + CH$_3$CHOH scheme. Both electronic (above) and zero-point corrected energies (below) are shown.}
  \label{fig:acac}
\end{figure*}

The reaction paths for both schemes are shown in Figures \ref{fig:glyc}, \ref{fig:acac}.

Figure \ref{fig:glyc} exhibits the reaction path following the  O($^3$P) + CH$_2$CH$_2$OH addition. The barrierless addition of oxygen leads to the (I) intermediate, which is about 400 kJ/mol more stable than the reactants. Its \it trans \rm counterpart, the slightly less stable (by $8$ kJ/mol) intermediate (II), can easily be reached from the \it cis \rm species through a $20$ kJ/mol barrier  (TS1). Both intermediates are then able to undergo dissociation to formaldehyde and CH$_2$OH through the transition states  (TS6) and (TS5). These dissociations exhibit barriers around $55$ kJ/mol. Other dissociations can also be observed from both these intermediates, leading to \it cis- \rm or \it trans-\rm glycolaldehyde and H, respectively through (TS2) and (TS3) that are about $115$ kJ/mol higher in energy than their corresponding dissociating intermediates. The most stable products that can be obtained with this path are formaldehyde + CH$_2$OH, with a relative energy of around $-346$ kJ/mol followed by the glycolaldehyde isomers, with a relative energy of around $-310$ kJ/mol. 

Wang and Bowie (2010) performed electronic structure calculations for the same global potential energy surface, but considering 
the possible formation of glycolaldehyde from the reaction H$_2$CO + CH$_2$OH. Their entrance channel, therefore, is one of the present exit channels 
in our case. The energy values of the corresponding intermediates and transition states can be compared and are in general good agreement. More specifically,
 our values for the energy of the transition states and global endothermicity are lower than the ones presented in that work. 
 Nevertheless, both are still high enough to render the reaction  H$_2$CO + CH$_2$OH $\rightarrow$ HCOCH$_2$OH + H
 prohibitive in the ISM.
 
Figure \ref{fig:acac} exhibits the reaction path following the O($^3$P) + CH$_3$CHOH addition. The barrierless addition of oxygen leads to the  (III) intermediate, which is about $410$ kJ/mol more stable than the reactants. This species is then able to undergo a dissociation into formic acid and the CH$_3$ radical, through the (TS11) transition state. This dissociation exhibits an approximate $45$ kJ/mol barrier. Another dissociation can also be observed, leading to acetic acid and H, through (TS9) which is about $60$ kJ/mol higher in energy than (III). Starting again from (III), elimination of an OH radical can occur (through the (TS8) transition state, lying approximately $100$ kJ/mol above the intermediate) yielding acetaldehyde. Furthermore, from (III) hydrogen migration can lead to compound  (IV) with a barrier of $105$  (TS7) kJ/mol. The intermediate (IV) is found to be around $460$ kJ/mol more stable than the precursor.  (IV) can also undergo hydrogen loss resulting in acetic acid and H through a $117$ kJ/mol barrier (TS10). The most stable products are formic acid and CH$_3$, with a relative energy of $-419.5$ kJ/mol with respect to the precursors, followed by acetic acid $+$ H that exhibits a relative energy of $-395.3$ kJ/mol.

\subsection{Kinetics calculations}
\label{sec:kinetic} 

As in previous work (Balucani et al. 2012, Leonori et al. 2013, Skouteris et al. 2015, 
Vazart et al. 2015, Sleiman et al. 2017) a combination of capture theory and 
the Rice-Ramsperger-Kassel-Marcus (RRKM) calculations was used
to determine the relevant rate coefficients and branching ratios. For the first steps
(addition of the O($^3$P) atom to either the CH$_2$CH$_2$OH or the
CH$_3$CHOH radicals) capture theory was used, whereas for the
subsequent reactions energy-dependent rate constants were calculated
using the RRKM scheme and taking into account anharmonicity of the
vibrational modes. Subsequently, the master equation was solved at all
relevant energies for both systems (to take into account the overall
reaction scheme), Boltzmann averaging was carried out to obtain
temperature-dependent rate coefficients and, finallly, rate coefficients
were fitted to the form $k = \alpha \times (T_{gas}/300K)^\beta \times
    exp[-\gamma/T_{gas}$. The
values of $\alpha, \beta$ and $\gamma$ in each case are given in Table \ref{tab:our-new-reacs}.

Back-dissociation is negligible in both cases, due to the high
stability of the initial intermediate and the presence of very exothermic channels
with low exit energy barriers.  When the O atom adds to
CH$_2$CH$_2$OH, the most probable fates of the radical are cleavage of
the C-C bond (to yield formaldehyde and the CH$_2$OH radical)
or an H atom elimination to give glycolaldehyde. The first one
dominates at all temperatures due to the significantly lower energy
barrier involved. Nevertheless, as can be seen from the final values,
a substantial percentage of the intermediate goes to glycolaldehyde (yield of 19\%).

An analogous situation presents itself when the O atom adds to the
CH$_3$CHOH radical. The two most probable fates of the radical are C-C
bond cleavage, yielding HCOOH and a methyl radical
 and elimination of an H atom from the $\alpha$ C atom to yield
acetic acid (CH$_3$COOH).  Before the elimination, there is also
the possibility of an H atom transfer from the $\alpha$ C atom to the
newly added oxygen, followed by an H atom elimination from one of the
two O atoms. However, this is the least followed path towards acetic
acid formation, both because of the higher barrier involved and the
longer reaction path. The most abundant product is formic acid
(with a yield of 59\%), because of the lower barrier involved, while acetic
acid formation is second highest (33.5\%). Finally, there is some possibility
of elimination of an OH radical from the initial intermediate,
yielding CH$_3$CHO. The barrier for this process is
considerably higher than both previous ones and therefore the rate of
formation of acetaldehyde is lower. Concerning this last point, it should be
noted that Edelb\"uttel-Einhaus et al. (1992) were, instead, only able to detect the
CH$_3$CHO product in their room temperature experiments.

\section{Astrochemical modelling}\label{sec:astrochemical-model}

\subsection{Description of the model}\label{sec:description-model}

In order to understand whether the proposed new reaction scheme and
rate coefficients can explain observations towards hot corinos, we used an
astrochemical model which simulates their conditions. Towards this scope,
we used a modified version (to improve its versatility) of the
time-dependent chemistry code NAHOON\footnote{The original code is
  publicly available at {\it http://kida.obs.u-bordeaux1.fr} (Wakelam
  2014).}. We run the code in two steps. In the first step, we
follow the chemical composition of the molecular cloud from which the
hot corino evolves. We start with the standard atomic state with the
element abundances listed in Table \ref{tab:ini-injected} and wait for
the steady state composition of the gas. We then simulate the hot
corino appearance by injecting into the gas phase the species
previously frozen into the dust grain mantles in the quantities listed
in Table \ref{tab:ini-injected}.  The assumption is that when the dust
grain temperature reaches $\sim$100 K the ice mantles sublimate and
all species trapped in the water matrix co-desorb with it. This is a
rough approximation, but enough for the scope of this article, the aim
of which is to provide an order of magnitude of the species
abundances.

 The starting point of our proposed chemical scheme is the sublimation of ethanol from interstellar grains. Since IR observations have (only) possibly identified this species in the ice mantles of interstellar grains (Boogert et al. 2015, Schutte et al., 1999; Oberg et al., 2011), its abundance is a parameter of the model, that we varied between $10^{-8}$ and $10^{-6}$.

Similarly, it is not clear what the abundance of atomic chlorine in
hot corinos is. On Earth, chlorine is mostly in oceans and very
little in rocks. Therefore, only a little fraction of
chlorine is probably contained in the refractory grains of the ISM (Jenkins et
al. 2009).
Observations of HCl in hot cores/corinos and shock sites show that
this molecule has an abundance $\sim 10^{-9}$ (e.g. Peng et al. 2010;
Codella et al.  2012; Kama et al. 2015), namely about 300 times lower
than the solar abundance. Since in hot cores/corinos and shock sites
the grain mantles components are injected into the gas phase, these
observations show that HCl is not the major reservoir of chlorine,
contrarily to model predictions (e.g. Neufeld et al. 2012). It is,
therefore, possible, if not likely, that a large fraction of Cl is
atomic. Since no observations exist to constrain the abundance of
atomic chlorine in hot corinos, its abundance is considered a
parameter of the model and it is varied between $10^{-9}$ and
$10^{-7}$. The highest value corresponds to the assumption that 70\%
of Cl is depleted in the refractory grains or in other Cl- bearing
molecules. 

The other crucial species involved in the initiating steps of the proposed
scheme is OH (Table \ref{tab:our-new-reacs}). This radical is a
product of the injection of water from the ice mantles and it is
self-consistently computed by the astrochemical model. In this case,
we adopted the ``standard'' value for injected water of $1\times 10^{-4}$
(e.g. Boogert et al. 2015), quoted in Table \ref{tab:ini-injected}.

Finally, the H density of the hot corinos is assumed to be
$2\times10^8$ cm$^{-3}$ and its temperature 100 K. The cosmic ray
ionisation rate is assumed to be $3\times10^{-16}$ s$^{-1}$
(e.g. Caselli \& Ceccarelli 2012).
\begin{table}
  \centering
  \begin{tabular}{lc}
    \hline \hline
    Species & Abundance \\
    \hline
    \multicolumn{2}{c}{Step 1: elemental abundances}\\
    He  & 9.0$\times10^{-2}$ \\
    O   & 2.6 $\times10^{-5}$ \\
    C   & 1.7$\times10^{-5}$ \\
    N   & 6.2$\times10^{-6}$ \\
    S   & 8.0$\times10^{-8}$ \\
    Si  & 8.0$\times10^{-9}$ \\
    Mg & 7.0$\times10^{-9}$ \\
    Fe  & 3.0$\times10^{-9}$ \\
    Na & 2.0$\times10^{-9}$ \\
    Cl  & 1.0$\times10^{-9}$ \\
    F   & 1.0$\times10^{-9}$ \\
    \hline
    \multicolumn{2}{c}{Step 2: injected mantle species}\\
    H$_2$O                & 1.0$\times10^{-4}$ \\
    CO                       & 2.0$\times10^{-5}$ \\
    CO$_2$                & 2.0$\times10^{-5}$ \\
    H$_2$CO              & 5.0 $\times10^{-6}$ \\
    CH$_3$OH            & 5.0 $\times10^{-6}$ \\
    NH$_3$                & 5.0 $\times10^{-6}$ \\
    CH$_4$                & 3.0 $\times10^{-6}$ \\
    CH$_3$CH$_2$OH & 1.0$\times10^{-8}$--1.0$\times10^{-6}$ \\
    Cl                        & 1.0$\times10^{-9}$--1.0$\times10^{-7}$ \\
    \hline
  \end{tabular}
  \caption{List of the species injected from the iced mantles (lower 
    half table), plus
    the elemental abundances of the molecular cloud phase (upper half table). The
    abundances of the injected species are
    taken from Boogert et al. (2015), who give them relative to
    H$_2$O. The elemental abundances are $5 \%$ the solar ones for oxygen, carbon and 
   nitrogen, and $0.5 \%$ for the heavier elements, to account for the freeze-out of these elements in the molecular cloud. Please note that the Cl abundance is a parameter of the model (see text). All abundances are with respect to H-atoms.}
\label{tab:ini-injected}
\end{table}

\subsection{Chemical network}\label{sec:chemical-network}
We used the KIDA network\footnote{The original network is publicly
  available at {\it http://kida.obs.u-bordeaux1.fr} (Wakelam et
  al. 2015).}, modified following Loison et al. (2014), Balucani et
al. (2015) and Skouteris et al. (2017), plus the reactions in Tables
\ref{tab:our-new-reacs} and \ref{tab:new-reacs}. The first table
reports the reactions in the new proposed scheme (Section
\ref{sec:new-reaction-scheme}), whereas the second table lists the
reactions added to complete the formation and destruction routes of
the newly introduced species (i.e. not present in the KIDA database) or
the reactions that were modified with respect to the KIDA content. 
Notes with the relevant references and arguments are listed in the
Appendix.

\begin{table*}
  \centering
  \begin{tabular}{lclccccl}
    \hline \hline
    \multicolumn{3}{c}{Reaction} & $\alpha$ & $\beta$ & $\gamma$ & Label & Notes\\
    \hline
    HCOCH$_2$OH + HX$^+$& $\rightarrow$&  HCOCH$_2$OH$_2^+$ + X & 2.0(-9) & 0 & 0  & 10 & 1 \\
    HCOCH$_2$OH + He$^+$& $\rightarrow$& HCO$^+$ + CH$_2$OH + He & 1.0(-9) & 0 & 0 & 11 & 2 \\
    HCOCH$_2$OH + H$^+$& $\rightarrow$& HCO$^+$ + CH$_2$OH + H    & 1.0(-9) & 0 & 0 & 12 & 2 \\
    CH$_3$COOH + HCO$^+$ &$\rightarrow$& CH$_3$CO$^+$ + CO + H$_2$O  & 2.5(-9) & 0 & 0 & 13 & 3 \\
    CH$_3$COOH + H$_3$$^+$ &$\rightarrow$& CH$_3$CO$^+$ + H$_2$ + H$_2$O   & 6.8(-9) & 0 & 0 & 14 & 3 \\
    CH$_3$COOH + H$_3$O$^+$ &$\rightarrow$& CH$_3$CO$^+$ + H$_2$O + H$_2$O  & 2.6(-9) & 0 & 0 & 15 & 3 \\
    CH$_3$COOH + H$^+$ &$\rightarrow$& CH$_3$CO$^+$ + H$_2$O  & 7.4(-9) & 0 & 0 & 16 & 3 \\
    CH$_3$COOH + He$^+$ &$\rightarrow$& CH$_3$CO$^+$ + OH + He  &  4.0(-9) & 0 & 0 & 17 & 3 \\
    CH$_2$CH$_2$OH /  CH$_3$CHOH + HX$^+$& $\rightarrow$&  CH$_3$CH$_2$OH$^+$ + X & 2.0(-9) & 0 & 0  & 18 & 4 \\
    CH$_2$CH$_2$OH / CH$_3$CHOH + H$^+$& $\rightarrow$&  CH$_3$CHOH$^+$ + H & 3.0(-9) & -0.5 & 0  & 19 & 5 \\
    CH$_2$CH$_2$OH / CH$_3$CHOH + He$^+$& $\rightarrow$&  C$_2$H$_4^+$ + OH + He & 3.0(-9) & -0.5 & 0  & 20 & 6 \\
    HCOCH$_2$OH$_2^+$ + $e$& $\rightarrow$& HCOCH$_2$OH + H & 1.5(-7) & -0.5 & 0 & 21 & 7 \\
    CH$_3$CHOH$^+$ + $e$& $\rightarrow$& H$_2$CO + CH$_3$ & 8.5(-7) & -0.74 & 0 & 22 & 8 \\   
    CH$_3$CHOH$^+$ + $e$& $\rightarrow$& H + H$_2$CO + CH$_2$ & 8.5(-7) & -0.74 & 0 & 23 & 8 \\   
    CH$_3$CHOH$^+$ + $e$& $\rightarrow$& H + HCO + CH$_3$ & 8.5(-7) & -0.74 & 0 & 24 & 8 \\
    CH$_3$CHOH$^+$ + $e$& $\rightarrow$& H + CO + CH$_4$ & 8.5(-7) &   -0.74 & 0 & 25 & 8 \\    
    CH$_3$CHOH$^+$ + $e$& $\rightarrow$& H + CH$_3$CHO & 3.0(-7) &   -0.74 & 0 & 26 & 8 \\    
\hline 
  \end{tabular}
  \caption{List of new reactions added to the chemical network. Notes
    on each reaction are reported in the Appendix.}
  \label{tab:new-reacs}
\end{table*}

We emphasise that the first step, which leads to CH$_3$CHOH
and CH$_2$CH$_2$OH
from ethanol, can be obtained either via the reaction with atomic Cl
or OH. Since in the literature there have been different values of branching ratios for the latter reaction, we considered 
two different scenarions 
(see Sec. \ref{sec:new-reaction-scheme}), namely the two sets of
reactions (3a), (4a), and (3b), (4b), respectively.

\subsection{Results}\label{sec:results}

We run three grids of models, each with the abundance of injected
ethanol and atomic chlorine in the range reported in Table
\ref{tab:ini-injected}. The three grids are obtained by varying the
conditions of the initiating steps of our proposed scheme (Fig.
\ref{fig:new-scheme}), namely the reactions of Table
\ref{tab:our-new-reacs} numbered 1 to 4.
In the first grid, we adopted the rates of (3a) and (4a), in the
second grid the rates of (3b) and (4b), and in the last grid we did
not consider the reactions (3) and (4), to quantify the role of atomic
chlorine. 

\begin{figure}
  \plotone{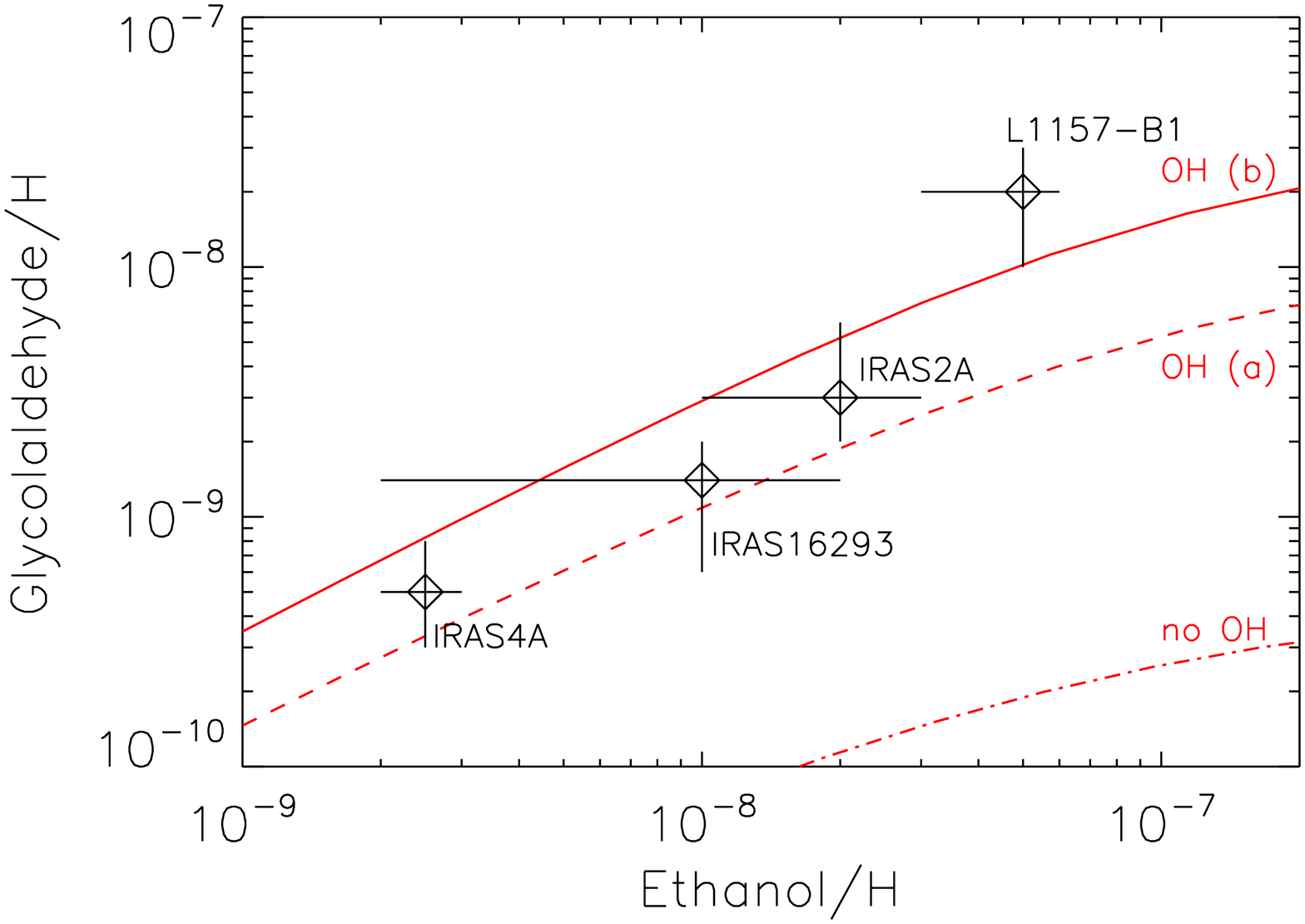}
  \caption{Abundance of glycolaldehyde as a function of ethanol
    abundance in the gas phase, which can be different from the one injected from the mantles. The computations refer to a gas with temperature equal
    to 100 K, H nuclei density $2\times 10^8$ cm$^{-3}$, cosmic ray
    ionisation rate $3\times 10^{-1}$ s$^{-1}$ and time
    $1.5\times 10^3$ yr. The three curves refer to models with
    different reactions of Table \ref{tab:our-new-reacs}: adopting the
    reactions (3b) and (4b) (solid line), reactions (3a) and (4a)
    (dashed line) and excluding the reaction with the OH radicals
    (dotted-dashed line). The atomic chlorine abundance is
    $2.2\times 10^{-8}$ in the computations.  Measured abundances
    towards NGC1333, IRAS4A and IRAS2A, IRAS16293-2422 and L1157-B1 are
    also reported with their uncertainties.}
  \label{fig:results-GA}
\end{figure}

The results at $1.5 \times 10^3$ yr (the approximate age of hot corinos
and of L1157-B1), are shown in Fig.  \ref{fig:results-GA}. The figure
shows the abundance of glycolaldehyde as a function of the ethanol
abundance in the gas, which can be different from the one injected
from the mantles (as it is used to make the other species). We predict
glycolaldehyde abundances in the range of $3\times 10^{-10}$ to
$2\times 10^{-8}$, and ethanol from $10^{-9}$ to $10^{-7}$.
The largest glycolaldehyde abundances are obtained adopting the most
favorable branching ratio of the reaction between ethanol and OH,
namely the rates (3b) and (4b) of Table \ref{tab:our-new-reacs}.
Ignoring the reactions (3) and (4) results in the lowest predicted
glycolaldehyde abundances. This means that the ethanol reaction with
Cl plays a minor role in our model, provided that a large abundance of water is present.

\begin{figure}
  \plotone{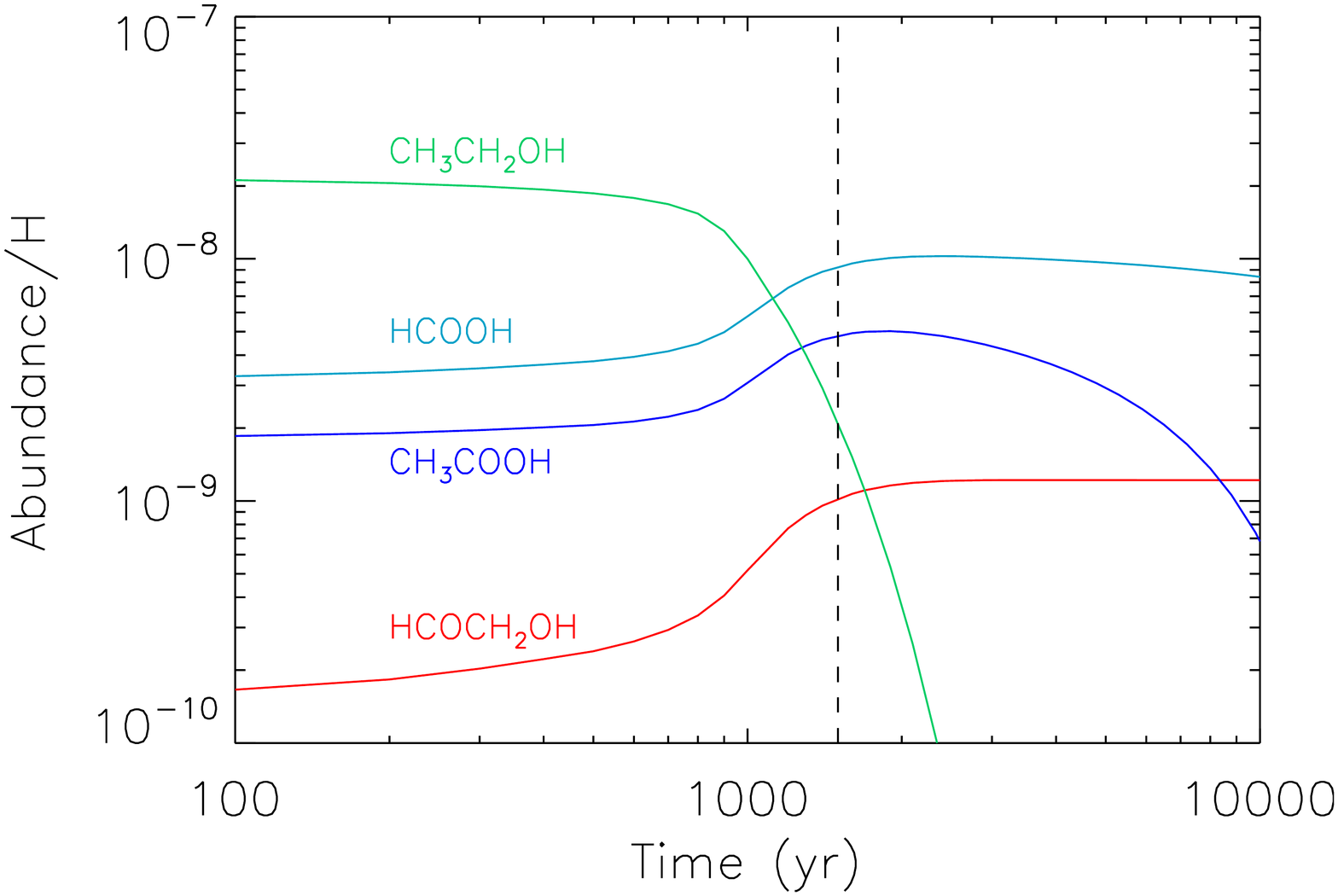}
  \caption{Abundance of glycolaldehyde (red),
    acetic acid (purple), formic acid (cyan) and ethanol
    (green) as a function of time. The computations are obtained
    for a gas with temperature equal to 100 K,
    H nuclei density $2\times 10^8$ cm$^{-3}$, cosmic ray ionisation
    rate $3\times 10^{-1}$ s$^{-1}$ and assuming reactions (3b) and
    (4b) of Table \ref{tab:our-new-reacs}. The abundance of injected
    ethanol and chlorine is $2.8\times10^{-8}$ and
    $2.2\times10^{-8}$, respectively. The black dashed line shows the
    time used to obtain Fig. \ref{fig:results-GA}, namely 1500 yr.}
  \label{fig:results-time}
\end{figure}

Figure \ref{fig:results-time} shows the abundance of glycolaldehyde,
acetic acid, formic acid and ethanol as a function of time. In these
computations, we adopted an abundance of injected ethanol and atomic
chlorine equal to $2.8\times 10^{-8}$ and $2.2\times 10^{-8}$,
respectively.  In the conditions assumed by the model (Section
\ref{sec:description-model}), the injected ethanol is all consumed in
about 2000 yr. Formic acid is the one that benefits most, followed by
acetic acid and, finally, glycolaldehyde. Before sublimated ethanol is fully
consumed, the abundance ratios are HCOOH/CH$_3$COOH$\sim$1.5 and
CH$_3$COOH/HCOCH$_2$OH$\sim$10, and are mostly governed by the
branching ratios of the first two steps of the proposed reactions
(Fig. \ref{fig:new-scheme} and Table \ref{tab:our-new-reacs}).
Once ethanol is fully consumed, the relative abundance ratios are
dominated by the destruction reactions (Table \ref{tab:new-reacs}).

\section{Discussion}\label{sec:disc-concl}

Table \ref{tab:obs} reports the measured abundances of the ethanol,
glycolaldehyde, acetic acid and formic acid in hot corinos (NGC1333
IRAS4A and IRAS2A, and IRAS16293-2422: Taquet et al. 2015; Coutens et
al. 2015; J\o rgensen et al. 2012; Jaber et al. 2014) and the shocked
site L1157-B1 (Lefloch et al. 2017)\footnote{These observations were
  obtained with the single-dish telescope IRAM-30m. Nonetheless, the
  abundances in Table \ref{tab:obs} were obtained via an accurate
  analysis taking into account a detailed knowledge of
  L1157-B1.}. Glycolaldehyde was detected also in two other solar type
protostars in NGC1333 (IRAS4B and SVS13A), with column densities
similar to the ones of IRAS4A and IRAS2A (De Simone et al. 2017), but
no measurements of ethanol, acetic acid and/or formic acid are
reported so far.

\begin{table*}
  \centering
  \begin{tabular}{lcccc}
    \hline \hline
    Source & \multicolumn{4}{c}{Abundance ($\times10^{-9}$)} \\
               & Ethanol & Glycolaldehyde & Acetic acid & Formic acid\\
     \hline
    NGC1333-IRAS4A & 2--3 & 0.2--6   &  &\\
    NGC1333-IRAS2A & 10--30     &    2--6   &  &\\
    IRAS16293-2422  & 2--20     & 0.6--2   & $\sim$ 0.2 & $\leq1$\\
    L1157-B1            & 30--60    &    10--30 & $\leq$2 & 7--10 \\
    \hline
  \end{tabular}
  \caption{Abundances, with respect to H-atoms, of ethanol,
    glycolaldehyde, acetic acid and formic acid towards three hot corinos, NGC1333
    IRAS4A and IRAS2A (Taquet et al. 2015), and IRAS16293-2422 (Jorgensen
    et al. 2016 and Jaber et al. 2014), and the shock site L1157-B1
    (Lefloch et al. 2017).}
  \label{tab:obs}
\end{table*}

The comparison of the model predictions with measurements of the
glycolaldehyde abundance is reported in Fig. \ref{fig:results-GA}.
When considering that the predictions shown in the figure are obtained
for generic hot corino conditions, the agreement with the observations
is very encouraging. Obviously, the model with the reactions set (3b),
(4b) produces a larger amount of glycolaldehyde (and a lower one of
formic acid and acetic acid).  In addition, the scheme that we propose
to synthesise glycolaldehyde from ethanol naturally explains the
correlation seen by Lefloch et al. (2017) between the abundances of
these two species.

On the contrary, the abundance of acetic acid and formic acid is
predicted to be about one order of magnitude larger than the ones
measured and reported in Table \ref {tab:obs}. We emphasise, however, 
that the only source in Table \ref {tab:obs} for which there is a measurement
of the formic acid abundance and an upper limit to that of acetic acid
is L1157-B1, which is not a hot corino, so that a more specific
modelling is necessary before firmly concluding that there is a problem.
Since recent observations of two other iCOMs, acetaldehyde and
formamide, in L1157-B1 show that there is segregation in their spatial
distribution, we postpone such a modelling to a dedicated forthcoming
article.

The other Table \ref {tab:obs} source with an estimate of the acetic
acid abundance and an upper limit to the formic acid one is
IRAS16293-2422. In this case, the discrepancy between the model
predictions and the observations might suggest that important routes
of destruction of the two species are missing in our network. Since
we carefully checked all the ``usual'' ion-neutral reactions of
destruction (i.e. with HCO$^+$, H$^+$, H$_3^+$, H$_3$O$^+$ and He$^+$), it is
possible that major sinks are due to missing reactions involving
abundant radicals. These reactions might possibly lead to an even
higher degree of molecular complexity following a scheme similar to the one proposed here.

\section{Conclusions}\label{sec:conclusions}

We have presented a new scheme for the synthesis of glycolaldehyde, acetic
acid and formic acid from reactions involving ethanol as an ancestor
species. 
The initiating reactions, with H-abstraction from
ethanol leading to two different reactive radicals, have been characterized
in laboratory experiments (see Section
\ref{sec:new-reaction-scheme}), even though further experimental work at the relevant temperature is
mandatory to determine the product branching ratios.
The subsequent reactions were considered here for the first time. As there was no information 
on the complete chemical scheme,
we have
performed dedicated electronic structure and kinetics calculations to
derive the rate coefficients and product branching ratios (Section
\ref{sec:comp-deta-results}).

The rate coefficients of the new reactions were inserted in an updated chemical network
and we ran several models to predict abundances of
glycolaldehyde, acetic acid and formic acid as a function of the abundance of
ethanol.
The predictions compare extremely well with the measured abundance of
glycolaldehyde in solar type hot corinos and shock sites, both in
terms of absolute abundance and in reproducing the correlation between
the ethanol and glycolaldehyde abundances observed by Lefloch et al.
(2017). Needless to say, more observations towards hot corinos
are mandatory to assess the robustness of our new network of reactions leading to glycolaldehyde.  The new observations and the detailed models of those sources will be able to discriminate whether the grain or gas-phase chemistry or a combination of the two are mainly responsible for glycolaldehyde formation.

On the contrary, acetic acid and formic acid are predicted to be about
ten times more abundant than the extremely sparse detections so far
available towards hot corinos and shock sites (only one in each case).
This might point to a lack of important routes of destruction of these
two molecules in our network, possibly via reactions involving radicals.  
Nonetheless, since observations are
published towards only two sources and the model presented here contain
a very generic description of hot corinos conditions, more
observations and source-dedicated modelling are necessary to confirm
this discrepancy. 

A more general conclusion is that the new gas-phase scheme
suggested in this article increases the number of studies that show the
important and previously overlooked role of neutral gas-phase chemistry in
the synthesis of iCOMs. First, since the detection of iCOMs in cold
prestellar objects (Bacmann et al. 2012, Cernicharo et al. 2012,
Vastel et al. 2014, Jimenez-Serra et al. 2016) it has been clear that
gas-phase reactions have to be relatively efficient at 10 K (Vasyunin
\& Herbst 2013, Balucani et al. 2015, Vasyunin et al. 2017) as the
so-called warm-up phase necessary in pure grain-surface models does
not take place in those objects. Second, new studies challenge the
exclusive role of grain-surface chemistry in the synthesis of iCOMs
also in the warm regions like hot corinos and shock sites (Barone et
al. 2015, Taquet et al. 2016, Skouteris et al. 2017, Codella et
al. 2017). The present study adds up new evidence that gas-phase
chemistry in the iCOMs synthesis has been overlooked, since important
reactions are missing in the current astrochemistry databases. 

However, the word ``end'' cannot be written yet, as more studies are
necessary. On the one hand, observations are too scarce to draw firm
conclusions and, on the other hand, more theoretical and experimental
studies are needed to complete the gas-phase networks. New experimental results on neutral-neutral reactions at low T, previously disregarded because of the presence of an entrance barrier, promise to boost the role of gas-phase reactions involving radicals (Potapov et al. 2017).

\section{Acknowledgements}

We acknowledge funding from the European Research Council (ERC) in the
framework of the ERC Advanced Grant Project DREAMS "Development of a
Research Environment for Advanced Modeling of Soft Matter", GA
N. 320951, and DOC ``The Dawn of Organic Chemistry'', GA N. 741002.
DS, NB, CeCe and ClCo thank Leonardo Testi, Eva Wirstr\"om and the 
Gothenburg Centre for Advanced Studies in Science and Technology for
the organization of the workshop Origins of Habitable Planets hosted
at Chalmers (Gothenburg) in May 2016, during which the basic ideas of
this work emerged.

NB acknowledges the financial support from the Universit\'e Grenoble
Alpes and the Observatoire de Grenoble.

This work has also been supported by MIUR "PRIN 2015" funds, project
"STARS in the CAOS (Simulation Tools for Astrochemical Reactivity and
Spectroscopy in the Cyberinfrastructure for Astrochemical Organic
Species)", Grant Number 2015F59J3R.


\software{Gaussian (Frisch et al. 2013), CFOUR (http://www.cfour.de), NAHOON (http://kida.obs.u-bordeaux1.fr)}

\appendix

\label{sec:appendix}

Note 1: To the best of our knowledge, proton transfer reactions by HCO$^+$, H$_3^+$ and H$_3$O$^+$ with glycolaldehyde have not been characterized in laboratory experiments. Nevertheless, Lawson et al. (2012), who have recently characterized electron-ion recombination of protonated glycolaldehyde, have reported that glycolaldehyde is easily protonated by H$_3^+$ without undergoing fragmentation (differently from acetic acid and methyl formate). We have therefore attributed a rate coefficient of $2 \cdot 10^{-9}$ cm$^3$ s$^{-1}$  (a rather typical value for proton transfer reactions involving large species) to the reaction with HX$^+$ where $X=$ CO, H$_2$, H$_2$O.

Note 2: To the best of our knowledge there are no laboratory characterizations of the reactions between He$^+$/H$^+$ with glycolaldehyde. Nevertheless, Cernuto et al. (2017) have recently characterized the reaction between He$^+$ and methyl formate, an isomer of glycolaldehyde. By far the main ionised fragment in their experiment was found to be HCO$^+$. Therefore, we have associated the HCO$^+$ formation channel to reactions with both He$^+$ and H$^+$. In addition, since Cernuto et al. have derived a rate coefficient much smaller than the Langevin value, we have employed here a rate coefficent of $1 \cdot 10^{-9}$ cm$^3$ s$^{-1}$.

Note 3: Proton transfer reactions by HX$^+$ (X = CO, H$_2$, H$_2$O) with acetic acid have been characterized in laboratory experiments. There have been some indications that, after proton transfer, the protonated acetic acid dissociates. In particular, Lawson et al. (2012) claimed that it was not possible to produce protonated acetic acid without dissociation. We follow their suggestion and associate the global rate coefficient (Anicich, 2003) to the formation of acetyl ion (CH$_3$CO$^+$) $+$ H$_2$O + X. As for the reaction with H$^+$ and He$^+$, we have also assumed that the main channel is the formation of acetyl ion and we have employed typical values of $\alpha$ for these processes. 

Note 4: In the absence of any data, we have assumed that the proton transfer from  HX$^+$ (X = CO, H$_2$, H$_2$O) is very effective for the CH$_2$CH$_2$OH/CH$_3$CHOH radicals (with a rate coefficient of $2 \cdot 10^{-9}$ cm$^3$ s$^{-1}$) and produces ionized ethanol (already present in the KIDA network).

Note 5: In the absence of any data, we have assumed that the interaction between CH$_2$CH$_2$OH/CH$_3$CHOH and H$^+$ causes a charge transfer producing protonated acetaldehyde. Since protonated acetaldehyde mostly recombines with an electron in a dissociative process, this assumption does not lead to significantly different results with respect to the choice of a dissociative charge transfer (see Note 8).

Note 6: In the absence of any data, we have employed a scheme similar to the He$^+$ reaction with ethanol with the same rate coefficient as in the UMIST database.

Note 7: In the absence of any data, we have assumed that electron recombination of protonated glycolaldehyde produces neutral glycolaldehyde with a typical rate coefficient for this kind of process. To be noted that significant fragmentation of glycolaldehyde can occur as already noted for other organic species (see, for instance, Hamberg et al. (2010), Vigren et al. (2013) and Geppert \& Larsson (2008)). Therefore, the amount of glycolaldehyde produced in the model can be considered an upper limit. Since a part of glycolaldehyde is recycled back into the proton transfer/electron recombination cycle, it is necessary to quantify this effect. We will do so in a future work.

Note 8: For the dissociative electron recombination of protonated acetaldehyde we have employed the Hamberg et al. (2010) values already present in the UMIST database.


\end{document}